\documentclass[pdflatex,sn-mathphys]{sn-jnl}

\usepackage{graphicx}
\usepackage{textcomp}
\usepackage{multicol}
\usepackage{xcolor}
\usepackage{placeins}
\usepackage{amsmath,amsfonts}

\jyear{2021}%

\theoremstyle{thmstyleone}%
\theoremstyle{thmstyletwo}%

\theoremstyle{thmstylethree}%

\begin{document}

\title[MNER]{Material Named Entity Recognition (MNER) for Knowledge-driven Materials Using Deep Learning Approach}


\author*[1]{\fnm{M. Saef Ullah } \sur{Miah}}\email{md.saefullah@gmail.com}
\author*[1,2]{\fnm{Junaida} \sur{Sulaiman}}\email{junaida@ump.edu.my}


\affil*[1]{\orgdiv{Faculty of Computing, College of Computing and Applied Sciences}, \orgname{Universiti Malaysia Pahang}, \orgaddress{\city{Pekan}, \postcode{26600}, \state{Pahang}, \country{Malaysia}}}
\affil*[2]{\orgdiv{Center for Data Science and Artificial Intelligence (Data Science Center)}, \orgname{Universiti Malaysia Pahang}, \orgaddress{\city{Pekan}, \postcode{26600}, \state{Pahang}, \country{Malaysia}}}




\abstract{The scientific literature contains a wealth of cutting-edge knowledge in the field of materials science, as well as useful data (e.g., numerical data from experimental results, material properties and structure). These data are critical for data-driven machine learning (ML) and deep learning (DL) methods to accelerate material discovery. Due to the large and growing amount of publications, it is difficult for humans to manually retrieve and retain this knowledge. In this context, we investigate a deep neural network model based on Bi-LSTM to retrieve knowledge from published scientific articles. The proposed deep neural network-based model achieves an f-1 score of \~97\% for the Material Named Entity Recognition (MNER) task. The study addresses motivation , relevant work, methodology, hyperparameters, and overall performance evaluation. The analysis provides insight into the results of the experiment and points to future directions for current research.}

\keywords{Named Entity Recognition, Material Named Entity Recognition, Materials Science, EDLC, Bi-LSTM}



\maketitle

\section{Introduction}

Material named entity recognition (MNER) is a task of named entity recognition (NER) in the field of Materials Science \cite{miah2022sentence}. NER stands for entity recognition or entity extraction. It is a natural language processing (NLP) technique that automatically detects and classifies named entities in a text. Individuals, organisations, locations, dates, amounts, monetary values, and percentages are all examples of entities. Entities in the text are considered as key information for the text which are important for understanding the context of the text. The term``Named Entity" and the word ``Named" are intended to limit the range of potential entities to those for which one or more rigid designators serve as the referent. When a designator designates the same item in all potential worlds, it is stiff. On the contrary, flaccid designators can refer to a variety of things in a variety of conceivable universes \cite{nadeau2007survey}. For example, in the field of material science, various material names, their property names, and synthesis processes can be defined as named entities for the field of material science. Therefore, it can be said that the task of named entity recognition for the field of material science can be called Material Named Entity Recognition.

The material named entity recognition model comprises of two steps like any other NER models which are, 
\begin{enumerate}
    \item Detection of a material named entity.
    \item Categorization of the detected named entity.
\end{enumerate}
The first step is to identify a word or series of words that together form an entity. Each word in a string represents a token. Each named entity can be formed from a single word or token or from a combination of words or tokens. For example, ``carbon" is a named entity with a single token; ``carbon monoxide" is a named entity with multiple tokens. This recognition can be done by a variety of methods, such as using rules, dictionaries, or machine learning \cite{goyal2018recent}.

The second phase, entails the establishment of material-specific entity types. For example, ``Mat," ``Proc," and ``Cmt" can all refer to different types of materials, synthesis processes, and characterization methods. These categories can be defined or generated by the expert from a variety of domains and as needed for a specific task.

As mentioned earlier, NER enables easy identification of key components within a text, and extraction of key components from a text enables organisation of unstructured data and detection of critical information; similarly, MNER is critical for dealing with large data sets or knowledge-based materials systems. The number of publications in materials science has increased by orders of magnitude over the last few decades. Now, a significant bottleneck in the materials discovery channel occurs when new results are compared to previously published literature. A possible solution to this problem would be to convert the unstructured raw text of published articles into structured database entries that are queryable programmatically. To accomplish this, text mining combined with material named entity recognition (MNER) task is performed to extract large amounts of information from the published materials science literature.

MNER is critical to data- or knowledge-driven materials science, also known as materials informatics \cite{jose2018materials}, which is an obvious component of the Industrial Revolution 4.0. MNER helps identify materials, synthesis processes, characterization methods, and many other types of entities that are essential for materials discovery research, identification of various synthesis processes to produce a substance or new object. MNER is the most important part of any knowledge-based materials system that deals with the discovery, extraction and knowledge representation of the discovered materials, processes and other entities from the published works.
The contributions of this study are as follows:
\begin{itemize}
    \item A deep neural network architecture for recognizing material and process entities from scientific articles.
    \item A comparison of the proposed model with different baseline machine learning models.
\end{itemize}
The rest of the paper is organized as follows, section \ref{s:relwork} presents relevant works, section \ref{s:method} presents the methodology employed in this study. Experiment and results are discussed in section \ref{s:result} and section \ref{s:conclusion} concludes the study.

\section{Related Work}
\label{s:relwork}

NER is undeniably a new field for the materials community. Training data is required for entity recognition models. If a domain already has knowledge bases, remote monitoring models can be used to train on known items and relationships. The most extensively used NER methods are dictionaries, rules, and machine learning, including deep learning. Knowledge-driven  Materials pipelines typically use all three methodologies. For efficient utilisation of annotated data, hybrid systems apply machine learning only when dictionaries or rules cannot manage the situation where the deep learning model process the data in sequential or semantic fashion. On the contrary dictionary searches cover material composition, chemical element names, properties, procedures, and experimental data.

A set of manually created rules or specifications that indicate how the relative order of rules and agreements should be handled is called a rule-based approach. Rules can be created using corpus-based methods, where multiple cases are analysed to find patterns, or by using domain knowledge and lexical conventions. 
LeadMine \cite{lowe2015leadmine} uses rules for naming conventions, ChemicalTagger \cite{chemicaltagger} analyses experimental synthesis sections of chemical texts, and parts of ChemDataExtractor use nested rules. For example, to use ChemDataExtractor \cite{Swain2016} with magnetic materials, researchers added domain-specific parsing rules with domain-specific terms (e.g., magnetic materials such as ferromagnetic and ferrimagnetic) \cite{court2018auto}.

Finally, machine learning-based algorithms identify specific entity names using a feature-based representation of the observed data. Since a sentence is a series of words, it is not sufficient to focus on the current word only. Sequential (and usually bidirectional) models that consider the preceding, the current, and the next word are required. Unlike rule-based approaches, supervised machine learning models require a huge amount of expert annotated data and strict annotation rules. Machine learning methods require careful evaluation of recognised classes and tag classification order. Kim et al.~\cite{kim2017materials}, Kononova et al.~\cite{Kononova2019}, Guha et al.~\cite{guha2021matscie}, and Weston et al. \cite{Weston2019} spearheaded NER work in the materials domain by implementing a bidirectional network of long and short term memory (LSTM) \cite{lstm} and a conditional random field (CRF) \cite{crf} for material entity identification. The Material NER problems differ by subarea. These include attributes, context, and reporting nuances. For example, Kononova et al. used a material parser to convert string representations of materials into chemical formulas, which were then split down into constituents and molar ratio balances. To find balanced reactions between precursor and target materials, the authors solved a system of equations. The open substances were generated from the precursor and target materials' combinations.

Because not all sorts of patterns can be implemented for all domains or a considerable amount of corpus data is required for a specific domain, the rule-based or dictionary-based approach is time-consuming and does not guarantee performance. After examining several research and works, this work eschewed these methodologies in favour of a machine learning-based strategy for the Material Named Entity Recognition task, which included both classic machine learning and deep learning approaches.

\section{Methodology}
\label{s:method}
\subsection{Problem Formulation}
The sequence labelling strategy is utilised for the entity recognition task. The sequence labelling job is processed using the form of the word, the context of the word in a sentence, and the word representation. Sentences ($S$) are tokenized from a piece of text ($T$), then tokens or words ($W$) are tokenized from the sentences, and finally each token is associated with a corresponding label ($L$). Formally, a piece of text $T$ contains a set of natural sentences $S$, $T = \{S_1,S_2,S_3,........,S_n\}$. Each sentence $S$ contains a sequence of $n$ tokens $S =  <W_1,W_2,W_3,........,W_n>$ and the corresponding labels $L = <l_1,l_2,l_3,........,l_n>$. The goal is to predict a list of tuples of tokens and associated labels $(W_i,l_i)$ from an input set of unknown entities.

\subsection{Deep Neural Network model for MNER Task}
Because this is a sequence labelling task, the Long Short Term Memory (LSTM) variation of the Recurrent Neural Network (RNN) \cite{RNN} is used. Since the RNN suffers from context difficulties as the sequence grows longer, the LSTM outperforms the original RNN. Along with the other layers of the deep neural network model, a bidirectional LSTM (Bi-LSTM) \cite{bi-lstm} layer is used in this study. Forward encoding of input tokens and reverse encoding of input tokens are combined in Bi-LSTM to provide the optimal context of a token inside a sentence. The forget gate, input gate, and output gate are the three gates in an LSTM network that update and regulate cell states. Hyperbolic tangent and sigmoid functions activate the gates. In response to incoming input data, the input gate controls how much new information will be encoded into the cell state. In reaction to new information entering the network, the forget gate evaluates whether information in the cell state should be removed. In the next time step, the output gate determines whether the information encoded in the cell state is supplied as input to the network.

\begin{figure*}[!ht]
	\centering
	\includegraphics[width=.6\linewidth] {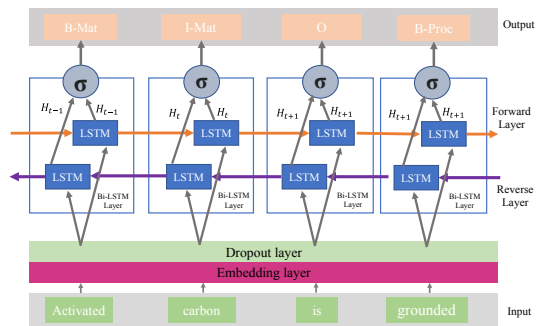}
	\caption{The proposed deep neural network model architecture for MNER task}
	\label{fig:propdnn}
\end{figure*}

The architecture of the deep neural network model developed for material named entity recognition task proposed in this study is shown in Figure \ref{fig:propdnn}.

The proposed deep neural network model has five layers, three of which are hidden layers and two of which are input and output layers, as shown in Figure \ref{fig:propdnn}. The first hidden layer is an embedding layer, which turns each word for a given sentence into a fixed-length vector. To use the embedding layer, all data is integer coded, which means that each word is represented by a distinct integer number. A spatial dropout layer is the following layer. To reduce model over-fitting and increase model performance, this dropout layer is employed as a regularisation strategy. This layer regularises the network during training by probabilistically removing the input and recurrent connections to the LSTM units from activation and weight updates. The bidirectional LSTM layer is put after the dropout layer. Two LSTM layers are introduced within the bidirectional Keras wrapper. The first LSTM model learns the provided sentence's word order, while the second LSTM model learns the reverse order of the first model. A time distributed wrapped dense layer is put after the LSTM layer. To maintain one-to-one relationships between input and output, this time distributed wrapper applies a layer to every temporal slice of an input. The proposed deep neural network model implementation can be expressed using the Algorithm \ref{alg:tm}.

The deep neural network is implemented using the Python programming language in combination with the Keras library and trained using the Tensorflow library.

\begin{algorithm}[!ht]
	\caption{Proposed Deep Neural Network model architecture}
	\label{alg:tm}
\begin{algorithmic}[1]
	\State Input: Sentence List with word and word-labels: $SL$ 
	\State $num\_words$ = unique words in dataset, $num\_tags$ = unique labels in dataset
	\State set $maxLen$ = 90
	\State For \{$S$ in $sent\_list$\}\{
			$X$ = pad sequences words
		\} 
		
	\State	For \{$S$ in $sent\_list$\}\{
			$y$ = pad sequences labels, $y$ = hot encode $y$ \}
		\State $x\_train$, $x\_test$, $y\_train$, $y\_test$ = train\_test\_split($X$, $y$, $test\_size$ = 0.1, $random\_state$ = 1)
		\State $input\_word$ = Input(shape = ($maxLen$,))
		\State $model$ = Embedding($input\_dim$ = $num\_words$, $output\_dim$ = $maxLen$, $input\_length$ = $maxLen$)($input\_word$)
		\State $model$ = SpatialDropout1D(0.2)($model$)
		\State $model$ = Bidirectional(LSTM($units$ = 200,$return\_sequences$ = True, $recurrent\_dropout$ = 0.2))($model$)
		\State $out$ = TimeDistributed(Dense($num\_tags$,$activation$ = `softmax'))($model$)
		\State $model$ = Model($input\_word$,$out$)
		\State $model$.compile($optimizer$ = `adam', $loss$ = `categorical\_crossentropy', $metrics$ = [accuracy, precision\_m, recall\_m, f1\_m])
		\State $model$.fit($x_train$, np.array($y\_train$), $batch\_size$ = 16, $verbose$ = 1, $epochs$ = 50, \State $validation\_split$ = 0.2, $callbacks$=[tensorboard\_cbk, es])
		\State $model$.save(`matrec.h5')
		\State Return $matrec.h5$
\end{algorithmic}
\end{algorithm}

\subsection{Evaluation Methods}
Precision, Recall, and $F1$ are used to evaluate the proposed Material Named Entity Recognition (MNER) model. When analysing entity predictions, if each token is correctly identified, the entity is marked as valid. When the entities are successfully predicted, True Positives ($TP$) are calculated; False Positives ($FP$) are marked when the predicted initial token does not match the entity's marked token. When the system forecasts the initial token of the predicted entities erroneously, False Negatives ($FN$) are registered. The equations stated in equation (\ref{eq:p}) are used to determine precision ($P$), recall ($R$), and the harmonic mean of precision and recall $F1$.

\begin{equation}
	\label{eq:p}
	P = \frac{TP}{TP+FP} \qquad
	R = \frac{TP}{TP+FN} \qquad
	F1 = \frac{2*P*R}{P+R}
\end{equation}

\section{Experiment and Results}
\label{s:result}
\subsection{Dataset}
In this study, a hand crafted dataset annotated by the domain expert from electric double layer capacitor (EDLC) is used \cite{matrec-data}. The dataset is curated from the full text of fifty scientific articles from EDLC domain. The text are annotated in Inside-Outside-Beginning (IOB)~\cite{IOB} format. There are five labelled classes in the dataset namely, $1.$ B-material, $2.$ I-material, $3.$ B-process, $4.$ I-process, and $5.$ O.
The summary of the dataset is presented in Table \ref{tab:dsoverview}.
\begin{table}[!ht]
	\centering
	\caption{Dataset overview}
	\label{tab:dsoverview}
	\begin{tabular}{ll}
		\toprule
		\textbf{Dataset Parameter}                                              & \textbf{Value} \\ \midrule
		Number of annotated article                                   & 50    \\ 
		Number of sentences having any entity                 & 1115  \\ 
		Number of annotated words                                      & 3155  \\ 
		Number of sentences containing Material entity                 & 980   \\ 
		Number of sentences containing Process entity                  & 265   \\ 
		Average number of sentences having any entity per article & 22.3  \\ 
		Average entity annotated per document                          & 63.1  \\ 
		Average entity annotated per sentence                          & 2.8   \\ 
		Average material entity containing word per document           & 51.1  \\ 
		Average process entity containing word per document            & 12    \\ 
		Average material entity containing word per sentence            & 2.6   \\ 
		Average process entity containing word per sentence            & 2.3   \\ \bottomrule
	\end{tabular}
\end{table}
\subsection{Hyperparameters}
The hyperparameter values for the different layers of the proposed deep neural network vary from layer to layer. For the LSTM layer, tanh is used as the activation function with 200 neurons and 0.2 is set as the recurrent dropout value. For the dense layer, softmax is used as the activation function. Adam optimizer is used as the optimizer and categorical\_crossentropy is used as the loss function. The length of the sequence is set to 60. The batch size is set to 16 when training the model and the model is trained using GPU. The proposed model uses an early stop-callback approach with minimal validation loss to avoid overtraining. To train the model, the network is fed 80\% of the sentences in the dataset and the model is tested on the remaining 20\% of the sentences with 5-fold cross-validation. The model is stored and evaluated after training and testing using various evaluation measures such as accuracy, precision, recall, and f1.

\subsection{Result Analysis}
The results of the proposed model compared to some renowned machine learning models are shown in Table \ref{tab:baseline_ml} in terms of precision, recall, and F1 scores.

\begin{table}[!ht]
	\centering
	\caption{Comparative results obtained from the experiment in terms of precision, recall and f1 score.}
	\label{tab:baseline_ml}
	
		\begin{tabular}{lccc}
			\toprule
			\textbf{Model Name}              & \textbf{Precision} & \textbf{Recall} & \textbf{F1}    \\ \midrule
			Proposed DNN model      & 0.965     & 0.957  & 0.961 \\ 
			DecisionTree Classifier & 0.948     & 0.9    & 0.938 \\ 
			ExtraTree Classifier    & 0.947     & 0.9    & 0.938 \\ 
			RandomForest Classifier & 0.945     & 0.88   & 0.927 \\ 
			XGBoost Classifier          & 0.93      & 0.74   & 0.839 \\ 
			Kneighbors Classifier   & 0.92      & 0.6    & 0.732 \\ 
			LGBM Classifier         & 0.91      & 0.37   & 0.526 \\ 
			Support Vector Classifier                     & 0.84      & 0.17   & 0.282 \\ 
			AdaBoost Classifier     & 0.83      & 0.17   & 0.282 \\ 
			Gaussian NB             & 0.82      & 0.09   & 0.162 \\ 
			Mimicking model \cite{guha2021matscie}          & 0.673     & 0.818  & 0.731 \\ \bottomrule
		\end{tabular}%
	
\end{table}

\begin{figure}[!ht]
	\centering
	\includegraphics[width=.8\linewidth] {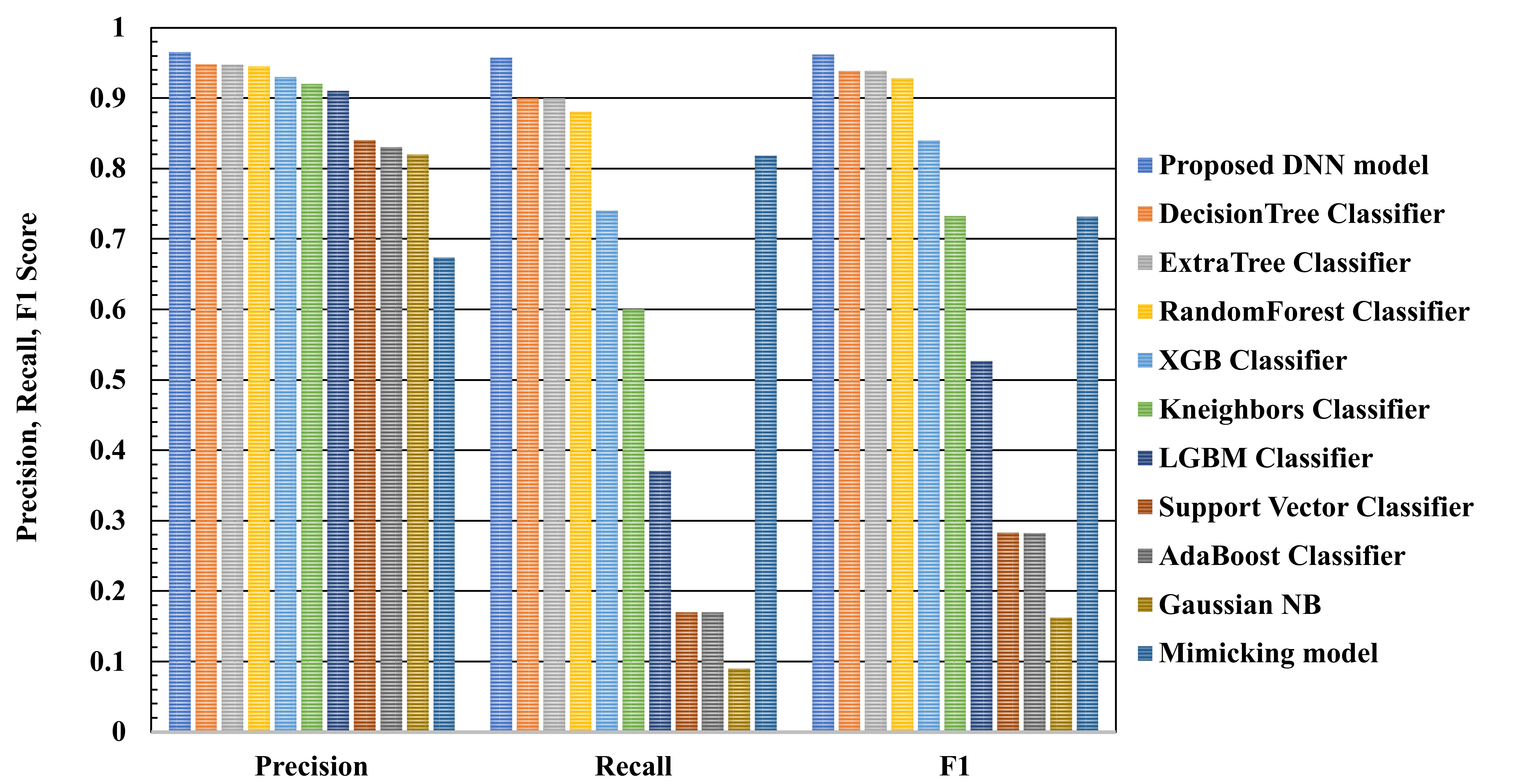}
	\caption{Precision, Recall and F1 comparison among different baseline machine learning models with proposed deep neural network model}
	\label{fig:prf-comp}
\end{figure}

Certain state-of-the-art baseline machine learning algorithms are compared to the proposed deep neural network model. The results obtained from these machine learning algorithms are presented in table \ref{tab:baseline_ml}. The findings of this experiment show that in the entity recognition task, the proposed deep neural network model outperforms many state-of-the-art machine learning models. The current result also shows that the proposed model performs better than the models using various pre-trained word embedding models and conditional random fields along with other layers of the network. The proposed DNN model achieved a better $f-1$ score than the system proposed by Guha et el. \cite{guha2021matscie}. The proposed model also achieved better $precision$ and $recall$ scores than the other compared models. The result shows that the proposed model is quite promising for the task of named entity recognition in the EDLC domain in terms of evaluation results. The comparison of the different evaluation metrics between the proposed model and other baseline models is shown in Figure \ref{fig:prf-comp}.

\section{Conclusion and Future Work}
\label{s:conclusion}
Overall, the research reported in this paper explores the possibilities of a deep neural network model based on intelligible LSTM in a knowledge-based material system. Our initial research focused on material entity recognition. Without significant knowledge of the material context, the deep neural network model performed admirably. We believe that this LSTM-based language model is a promising direction toward a more sophisticated NLP system for extracting material knowledge from scientific literature, since deep learning-based models are designed to adapt to a variety of different NLP tasks rather than focusing on a single task.


\bibliography{sn-bibliography}


\end{document}